\documentclass[aps,prl]{revtex4}

\usepackage{graphicx}
\begin{document}
\title{A global simulation for laser driven MeV electrons in $50\mu m$-diameter fast ignition targets}
\author{C. Ren$^1$, M. Tzoufras$^2$, J. Tonge$^3$, W. B. Mori$^{2,3}$, F. S. Tsung$^3$, M. Fiore$^4$, R. A. Fonseca$^{5,4}$,  L. O. Silva$^4$, J.-C.  Adam$^6$, and A. Heron$^6$}
\address{1. Department of Mechanical Engineering,  Department of Physics and Astronomy,  and Laboratory for Laser Energetics, University of Rochester, Rochester, NY 14627. \\
2. Department of Electrical Engineering, University of California, Los Angeles, CA 90095. \\
3. Department of Physics \& Astronomy, University of California, Los Angeles, CA 90095. \\ 
4. GoLP/Centro de Fisica dos Plasmas, Instituto Superior Tecnico, 1049-001 Lisboa, Portugal.  \\
5. DCTI, Instituto Superior de Cincias do Trabalho e da Empresa, 1649-026, Lisboa, Portugal.\\
6. Ecole Polytechnique, France.
}
\date{\today}
\begin{abstract}
The results from 2.5-dimensional Particle-in-Cell simulations for the interaction of a picosecond-long ignition laser pulse with a plasma pellet of 50-$\mu m$ diameter and 40 critical density are presented. The high density pellet is surrounded by an underdense corona and is isolated by a vacuum region from the simulation box boundary.  The laser pulse is shown to filament and create density channels on the laser-plasma interface. The density channels increase the laser absorption efficiency and help generate an energetic electron distribution with a large angular spread. The combined distribution of the forward-going energetic electrons and the induced return electrons is marginally unstable to the current filament instability. The ions play an important role in neutralizing the space charges induced by the the temperature disparity between different electron groups. No global coalescing of the current filaments resulted from the instability is observed, consistent with the observed large angular spread of the energetic electrons.
\end{abstract}
\pacs{} 
\maketitle

\narrowtext
\section{introduction}
\label{intro}
The fast ignition concept, first proposed more than a decade ago\cite{tabak:pop:fi},  provides a possible alternative path to achieve ignition in inertial confinement fusion with the potential of reducing drivers energy and increasing gains significantly. In this concept, the ignition is first achieved in a small region ($L\sim20\mu m$) within a compressed pellet by a $\sim$10 ps intense flux of MeV electrons. The MeV electrons are produced by the absorption of a petawatt (PW)-class laser at the edge of the pellet. Recent experimental results in Japan\cite{kodama:nature} showed 3 orders of magnitude increase of neutron yield with a 0.5-PW ignition laser and consequently spurred intense new research activities in this area. One of the research foci is to predict the ignition laser energy needed to ignite a certain target. This energy is determined by a number of highly nonlinear and dynamic processes such as the laser propagation in the underdense corona,  the laser hole-boring and electron heating near the critical surface,  the transport of the energetic electrons in the overdense plasma, and the subsequent heating of the ignition region by these electrons. No theory exists and simulations and experiments are the only tools to study these processes. 

\begin{figure}[!th]
\begin{center}
\includegraphics{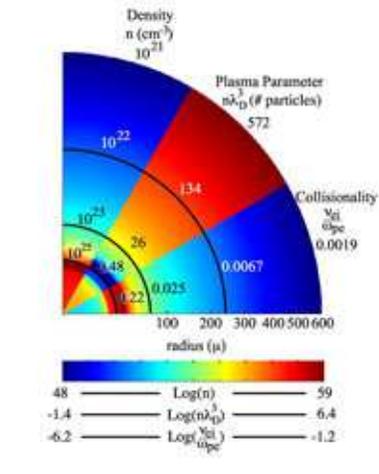}
\caption{(color online) Electron density and temperature (in terms of $n\lambda_D^3$ and $\nu_{ei}/\omega_{pe})$ profiles of a typical target at its peak compression. } 
\label{fig1}
\end{center}
\end{figure}

Particle-in-Cell (PIC) simulations provide a first-principle, detailed description for the plasma region of the pellet, which comprises the majority of the pellet except for the very dense core (electron density $n_e>10^{23}/cm^3$). Since the PIC simulations are computationally intensive, currently it is not possible to perform full scale simulations for the entire pellet. Figure \ref{fig1} shows the $n_e$ and electron temperature ($T_e$) information of a cryogenic target, imploded by a 25-kJ laser, at its peak compression\cite{betti}. The $T_e$ information is represented by the plasma parameter (number of electrons in a Debye cube, $n\lambda_D^3$) and the ratio of the electron-ion collisionality to the electron plasma frequency, $\nu_{ei}/\omega_{pe}$. In the region where $n_e<10^{23}/cm^3$, $n_e\lambda_D^3>>1$ which satisfies the usual definition of a plasma. Notice here $n_e\lambda_D^3$ is not overly large either, which indicates that the discrete particle effects are important. In this region $\nu_{ei}/\omega_{pe}<<1$, which shows that plasma behavior is not dominated by the collisions. All these make the PIC model the ideal simulation tool for this region. However, even if we were to simulate a plasma slab of 250 $\mu m$-long and $50\mu m \times 50 \mu m$ in cross section and resolve the smallest relevant physical length $c/\omega_{pe}$ ($\Delta x=1/3c/\omega_{pe}$) for $n_e=5\times10^{22}/cm^3$, the number of cells needed would be $1.3\times 10^{12}$. Assuming 4 particles/cell for each species (electrons and ions), the number of particles needed would be $10^{13}$. The total memory required to store the particle momentum and position information would be 560 Tera-Bytes (TB). Assuming a time step limited by Courant condition of $\Delta t=1/(3\sqrt{3})\omega_{pe}^{-1}$ and total simulation time of 10 ps, the total number of steps needed would be $6.7\times 10^{5}$. The total number of floating-point operations (FLOPs) would be $2.7\times 10^{20}$ (assuming 40 FLOPs/particle-step). This would take 10 months on a computer capable of 10 Tera FLOPs/second (TFLOPS). This kind of simulations can be routinely performed only when 100-1000 TFLOPS computers with large memory are widely available. Most recent fast ignition PIC simulations used scaled-down targets with reduced sizes and/or in two-dimension (2D) instead of three-dimension (3D)\cite{wilks,pukhov:prl:fi,lasinski:pop:fi,sentoku:pop:00,ren:prl:fi}. Many of the important issues, such as hole-boring, laser absorption, and current filamentation, have been studied.

However, some key physics depends critically on global electron paths in the target and thus depends on the target size and the boundary conditions used in the simulations. For example, to absorb 30\% of the energy in a $10^{20} W/cm^2$ laser for 1 ps, the thickness  of a plasma slab with $n_e=10^{22}/cm^3$ needs to be $190\mu m$, assuming every electron in the slab leaves with an energy increase of 1 MeV. However, the laser cannot sweep this distance in 1 ps through hole-boring. Even if it could, the ion layer formed from electron depletion would pull the electrons back. The actual absorption is achieved through continuous replenishing of cold electrons from the bulk of the target into a narrow region in contact with the laser front. If the simulated target is too small and cannot provide enough needed cold electrons, the MeV electrons may not leave the interaction region and may be heated furthermore to achieve a different distribution.

\begin{figure}[!th]
\begin{center}
\includegraphics{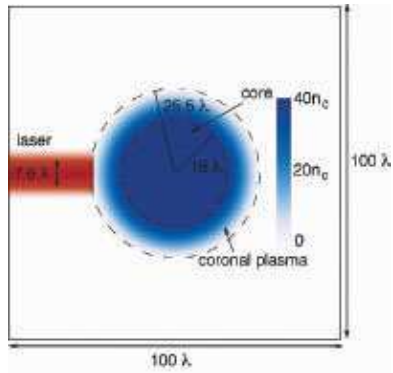}
\caption{(color online) 2D PIC simulation setup. } 
\label{fig2}
\end{center}
\end{figure}

In this paper we present results from a series of 2.5D (with two space dimensions but all three components of particle velocities and electromagnetic fields) PIC simulations with the PIC code OSIRIS\cite{osiris} in a simulation box of $100\lambda\times100\lambda$ with a round target of $51\lambda$ diameter. ($\lambda$ is the laser wavelength and also the basic length unit the simulations are scaled to.) The setup (Fig.\ref{fig2}) allows the simulation of the plasma-vacuum boundary which provides an important alternative pathway for the MeV electrons. The target consists of a proton-electron plasma with $T_e=7.4$ keV and $T_i=1$ keV. The target has a uniform density core with $n_e=40n_c$ within a diameter of $32\lambda$ and a coronal ring with the density linearly decreasing from $40n_c$ to 0 in 9.5$\lambda$. Here $n_c$ is the critical density which corresponds to $1.1\times 10^{21}cm^{-3}$ for $\lambda=1\mu m$. The laser is launched from the left boundary with peak intensities of $I=10^{20}$ and $10^{21} (1\mu m/\lambda)^2 W/cm^2$ and a rise time of $19$ laser period, after which the laser amplitude is kept constant. The laser
transverse profile is a Gaussian with a diffraction limited spot size of $w=7.5\lambda$. Both s- and p-polarization are used to infer 3D effects. A typical simulation duration is 309 laser period,  which corresponds to about 1 ps for $\lambda=1\mu m$.

To resolve the skin depth for $n=40n_c$, $c/\omega_{pe}\approx\lambda/40$, we use a grid size of $\Delta x\omega_{pe}/c=\Delta y\omega_{pe}/c=0.33$ with $12032\times12032$ grid cells. Therefore  $\Delta x\approx3\lambda_D$. We reduce numerical self-heating\cite{birdsall}  by smoothing the current. The time step used is $\Delta t\omega_{pe}=0.23$. We use 4 particles each for the electrons and the protons ($2.4\times10^8$ total particles) and their charges are weighted according to the initial local density.  The effective finite size particle collision frequency is roughly $\nu_{ei}/\omega_{pe}\approx0.035$\cite{hockney:jcp}. The particles are reemitted with their initial temperatures when they hit the $x$-boundary and are subject to periodic boundary condition in the $y$-direction. The boundary conditions for the electromagnetic fields are open in the laser propagation direction ($x$) and periodic in the transverse direction ($y$). A typical run consists of $1.6\times10^{13}$ particle-steps. 

The rest of the paper is devoted mainly to two issues: the laser absorption and electron heating near the laser-plasma interface (Sec.\ref{II}) and the filamentation of the electron flux inside the target (Sec.\ref{III}). A discussion and a summary are provided in Sec.\ref{IV}.

\section{Laser absorption and electron heating}
\label{II}
\begin{figure}[!th]
\begin{center}
\includegraphics{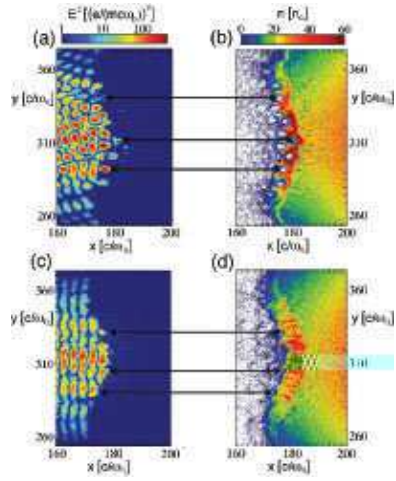}
\caption{(color online) Laser intensity $E_y^2$ (left column, in unit of $(mc\omega/e)^2$) and electron charge density (right column, in unit of $n_c$) for the s- (top row) and p-polarization (bottom row) cases at $t=648/\omega$. Here $\omega$ is the laser frequency and distance is in unit of $c/\omega$.} 
\label{fig3}
\end{center}
\end{figure}

As the laser propagates through the short underdense region here and reaches the critical surface, the most prominent feature observed from the simulations is the filamentation of the laser accompanied by the rippling of the target surface  (Fig.\ref{fig3}). The observed filament width is roughly an optical wavelength. A laser can filament due to either relativistic mass or ponderomotive force effects.  For the parameters of the fast ignition, both mechanisms will occur on timescales that are very short compared to the laser pulse length. Therefore, it is not as important to study the initial growth phase as it is to understand the eventual nonlinear steady state. The non-relativistic theory of ponderomotive filamentation predicts that the mode number with the fastest growth rate increases with plasma density and the growth rate is higher in the direction perpendicular to the polarization than in the polarization direction\cite{kaw}. These conclusions are verified in Fig.\ref{fig3} which shows that the s-polarized cases (laser electric field $E$ out of the simulation plane) display stronger laser filamentation than the p-polarized cases ($E$ in the simulation plane). With their higher intensity the laser filaments dig density channels at the laser-plasma interface via the ponderomotive force (Fig.\ref{fig3}, right column). These micro density channels in turn focus the laser filaments like converging waveguides which dig deeper channels. Therefore, the laser-plasma interface is unstable to any initial transverse modulation of the laser intensity, which can be viewed as filamentation in the overdense plasma\cite{valeo}. 

\begin{figure}[!th]
\begin{center}
\includegraphics{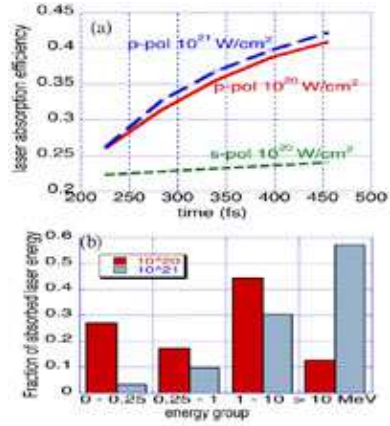}
\caption{(color online) Fraction of laser energy absorbed by the electrons for different laser intensities and polarizations (a) and the electron energy composition in the p-polarization case for two laser intensities at t=510 fs (b). All laser intensities and time are calculated assuming the laser wavelength is 1 $\mu m$.} 
\label{fig4}
\end{center}
\end{figure}

The density channels on the laser-plasma interface can enhance the laser absorption significantly. Heating mechanisms for normal-incident lasers include the normal coupling of the laser oscillation and the plasma wave at the critical surface\cite{anom_res_kaw} and the $J\times B$ heating\cite{jxb_kruer}. The density channels provide locations where the laser $E$-field can be in the density gradient direction. This provides additional heating mechanisms such as the enhanced resonant heating\cite{res_heating_kru} and Brunel heating\cite{brunel}. In our 2.5D simulations, the additional heating mechanisms are present only for the p-polarization cases. Therefore, we expect that the laser absoprtion is larger in the p-cases than in the s-cases. Figure \ref{fig4}a shows the laser absorption rate for the electrons, defined as the ratio of the electron kinetic energy increase to the incoming laser energy within a certain time interval $\Delta t$, $\eta (t)=[KE(t+\Delta t)-KE(t)]/(P_L\Delta t)$, for both s- and p-polarizations and verifies this conclusion. In addition, the p-cases also show that $\eta$ increases with time, consistent with the fact that as the channels becomes deeper these additional heating mechanisms become more efficient. On the contrary $\eta$ in the s-case remains almost flat. The absorption does not change significantly when the laser intensity changes from $10^{20} W/cm^2$ to $10^{21} W/cm^2$. However, the electron energy composition changes significantly (Fig.\ref{fig4}b). In the $I=10^{21} W/cm^2$ case, more than half of the absorbed laser energy goes to the electrons with energy greater than 10 MeV. These super-hot electrons may be too energetic to be stopped in the target core and be useful to the fast ignition. The electron spectrum in the $I=10^{20} W/cm^2$ cases can be approximated by a power law, which begins at $E\sim0.2$ MeV and falls off as $E^{-(2-3)}$\cite{ren:prl:fi}.

\begin{figure}[!th]
\begin{center}
\includegraphics{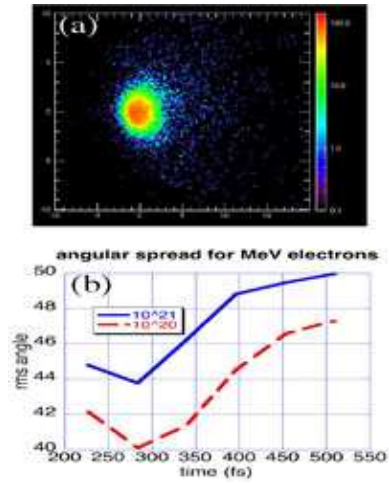}
\caption{(color online) (a) The $P_xP_y$ (in unit of $mc$) phase space of the electrons localized near the laser-plasma interface for the $I=10^{20} W/cm^2$ and p-polarization case (t=454 fs). (b) The rms angle (in degree) of those electrons with energy above 1 MeV for two laser intensities (both p-polarization). All laser intensities and time are calculated assuming the laser wavelength is 1 $\mu m$.} 
\label{fig5}
\end{center}
\end{figure}

The additional heating mechanisms\cite{res_heating_kru,brunel} allow the electrons to be accelerated directly in the laser $E$-field direction. Thus the energetic electrons are expected to to have a large transverse momentum ($P_y$). This can be seen from the $P_xP_y$ phase space plot for the electrons localized in front of the laser (Fig.\ref{fig5}a). It is also clear from the phase space plot that the energetic electrons do not form a beam with narrow emittance. The root-mean-square angle in the $xy$-plane for the above-1 MeV electrons is plotted in Fig.\ref{fig5}b, which also shows the large angular spread for both $10^{20} W/cm^2$ and $10^{21} W/cm^2$ cases. The large angular spread is one of the most important characteristics of the energetic electron distribution in these simulations and must be taken into account in studying their subsequent transport and in accessing the fast ignition feasibility.

\section{Current filament instability}\label{III}
If left alone, the laser-generated energetic electrons would deposit their energy over a large area when they reach the dense core region due to their large angular spread. However, the transport of the energetic electrons is not a simple free-streaming process but a highly nonlinear one during which their interactions with emerging magnetic fields play a very important role. For example, we have observed that the laser-heated electrons can generate intense magnetic fields through the $\nabla T\times \nabla n$ effect\cite{gradnt} on the target surface (not shown here). These magnetic fields, together with the surface radial electric fields, cause a fraction of the energetic electrons to move along the surface through $E\times B$-drift, which actually prevents the electrons from hitting the simulation box boundary. The motion of the energetic electrons that are shot  forward into the interior of the target is also influenced by magnetic fields, only in this case it is the current filament instability that provides these fields.

To properly study the current filament instability in the fast ignition one needs to add space charge effects due to temperature disparity between different groups of electrons to the standard analysis of Weibel instability\cite{weibel:prl,davidson:pf:72}. The $\sim$100 MA current carried by the MeV electrons in the fast ignition greatly exceeds the Alfven current limit\cite{alfven:pr:39},  which is set by the pinching force on the beam electrons from the self-magnetic field and is $\sim$30 kA for the MeV electrons.  A return current in the background plasma must be induced to neutralize the magnetic field of the beam and render the whole system nearly current neutral. Unlike the previously studied system of two identical counter-propagating electron clouds\cite{davidson:pf:72}, the forward energetic electrons have a much higher transverse temperature than the background plasma electrons carrying the return current.  Therefore, the forward current and return current pinch to different degrees when filamenting and a space charge imbalance will develop. The electric field from the space charge is along the wave vector  (or the $y-$direction here) and resists further filamenting. The space charge effects are especially important when the system is in the marginally unstable regime which is the typical case for the fast ignition simulations here. Furthermore, the ions will respond to the electric field and their motion becomes important in the marginally unstable case where the growth rate is comparable to the ion plasma frequency\cite{ren:prl:fi,honda:pop:00}. With a totally different origin from that in Ref.\cite{bret}, the space charge effects discussed here exist even when the system is stable to the two-stream instability.

The detailed analysis of the current filament instability with the space charge effects is presented in a separated paper\cite{tzoufras:prl}. Here we illustrate the basic space charge effects on the filament instability by considering a plasma system with 3 species: one species of hot energetic electrons moving in the $+{\hat x}$-direction with a drifting velocity $V_{d1}$, another species of cool return electrons moving in the $-{\hat x}$-direction with a drifting velocity $V_{d2}$, and a cold ion species. Specifically, the electron equilibrium distribution functions are assumed to be $f_{l0}({\bf v}, {\bf x})=n_{l}/(2\pi\sqrt{v_{tyl}v_{txl}})\exp(-v_{y}^{2}/2v_{tyl}^{2})\exp[-(v_x-V_{dl})^2/2v_{txl}^{2}]$, where $n_{l}$ is the density and $v_{txl}$ and  $v_{tyl}$ are the thermal velocities in the ${\hat x}$- and  ${\hat y}$-directions for each electron species, $(l=1,2)$. Under the relevant geometry and polarization $v_z$ is a negligible variable. The ion equilibrium distribution function is $f_{30}=n_{3}\delta(v_x)\delta(v_y)$. Both charge and current densities are zero in the equilibrium, $\sum_{l=1}^{3}q_{l}n_{l}=0$ and $n_{1}V_{d1}+n_{2}V_{d2}=0$. Here $q_1=q_2=-e$ are the electron charge and $q_3$ the ion charge. Consequently, there is no equilibrium electric or magnetic field.

To study the linear stability of this system against the filament mode, we assume that the mode propagation vector is in the $y$-direction, ${\bf k}=k {\hat y}$, and all perturbations have spatial and temporal dependence of $\exp(\gamma t- iky)$. For each species, the perturbed distribution function $f_{l1}$ can be found from the linearized Vlasov equation,
\begin{equation}
f_{l1}=i{(q_{l}/m_{l})\over \omega -kv_y}({\bf E}+{{\bf v}\over c}{\bf \times B})\cdot{\partial f_{l0}\over\partial{\bf v}},\label{f1}
\end{equation}
where $m_{l}$ is the mass of the particle in the $l$-th species. The relevant perturbed fields are ${\bf E}=E_x{\hat x}+E_y{\hat y}$ and ${\bf B}=B_z{\hat z}$. These can be related to the charge density $\rho$ and current density $j_x$ using the Maxwell's equations.
The linear analysis yields the following two coupled equations,
\begin{eqnarray}
\rho\{1+\sum_{l=1}^{2}{\omega_{pl}^2\over k^2 v_{tyl}^2}[1+\xi_{l}Z(\xi_{l})]-{\omega_{p3}^2\over\omega^2}\}+{j_{x}\over c}\sum_{l=1}^{2}{\omega_{pl}^2 V_{dl}c\over(\omega^2- k^2c^2) v_{tyl}^2}[1+\xi_{l}Z(\xi_{l})]=0,\label{eq1xi}\\
\rho\sum_{l=1}^{2}{\omega_{pl}^2 V_{dl}\over k^2v_{tyl}^2c}[1+\xi_{l}Z(\xi_{l})]+{j_{x}\over c}\{1+\sum_{l=1}^{2}{\omega_{pl}^2(V_{dl}^2+v_{txl}^2)\over (\omega^2-k^2 c^2)v_{tyl}^2}[1+\xi_{l}Z(\xi_{l})]-\sum_{l=1}^{3}{\omega_{pl}^2\over\omega^2-k^2 c^2}\}=0,\label{eq2xi}
\end{eqnarray}
where $\omega_{pl}\equiv\sqrt{4\pi q_{l}^2 n_{l}/m_l}$ is the plasma frequency of each species, $\xi_l\equiv\omega/(\sqrt{2}kv_{tyl})$, and $Z$ is the plasma dispersion function\cite{fried:z}.

The analysis of the dispersion relation from Eqs.\ref{eq1xi}-\ref{eq2xi} reveals two regimes of instability\cite{tzoufras:prl}. In the relative fast growth regime, $(m/M)kv_{ty2}<<\gamma<<kv_{ty2}$, the ions can be treated as immobile (ion mass $M\rightarrow\infty$ and ion plasma frequency $\omega_{pI}=0$) and the instability threshold is
\begin{eqnarray}
\label{threshold_f}
\sum_{l=1}^2\omega_{pl}^2{V_{dl}^2+v_{txl}^2\over v_{tyl}^2}>k^2c^2+\sum_{l=1}^2\omega_{pl}^2+{(\sum_{l=1}^2{\omega_{pl}^2V_{dl}\over v_{tyl}^2})^2
\over k^2+\sum_{l=1}^2{\omega_{pl}^2\over v_{tyl}^2}}.
\end{eqnarray}
The third term on the right hand side of Eq.\ref{threshold_f} comes from the off-diagonal term in Eqs.\ref{eq1xi}-\ref{eq2xi} that couples the perturbations in $\rho$ and $j_x$ and therefore represents the space charge effects. This term is positive definite, indicating that the space charges always raise the instability threshold. If the two electron species have the same transverse temperature, $v_{ty1}=v_{ty2}$, the space charge term vanishes since the current neutrality condition implies $\sum \omega_{pl}^2V_{dl}=0$. Therefore, the space charges originate from the temperature disparity between the two electron groups. In general, the space charge term decreases the growth rate and reduces the range of $k$ of the instability\cite{tzoufras:prl}.

In the slow growth regime, $\gamma<<(m/M)kv_{ty2}$, the ion response must be considered. Then the instability threshold turns out to be
\begin{eqnarray}
\label{threshold_s}
\sum_{l=1}^2\omega_{pl}^2{V_{dl}^2+v_{txl}^2\over v_{tyl}^2}>k^2c^2+\sum_{l=1}^2\omega_{pl}^2+\omega_{pI}^2.
\end{eqnarray}
Compared to Eq.\ref{threshold_f}, we can see in this marginal instability regime, the growth rate is so small that the ions always have time to react to cancel any potential space charge. Therefore, the instability threshold is the same as that when no space charge effect is considered\cite{davidson:pf:72}. 

\begin{figure}[!th]
\begin{center}
\includegraphics{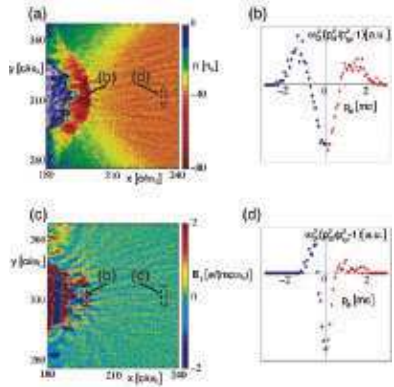}
\caption{(color online) (a) The electron charge density (in unit of $n_c$) at $t=1080 c/\omega$. The two boxes indicate where the stability analysis in (c) and (d) are carried out. (b) The magnetic field $B_z$ in unit of $mc\omega/e$, which is 107 MG for $1\mu m$-laser, at $t=1296 c/\omega$. The distances in (a) and (b) are in unit of $c/\omega$. (c) The current filament stability analysis for the shock region (indicated by the left box in (a)). (d) The same stability analysis for the target interior (indicated by the left box in (a)). A dot above the horizontal axis indicates its contribution to the instability. See the text for details. The case shown here is with $I=10^{20} W/cm^2$ and p-polarization.} 
\label{fig6}
\end{center}
\end{figure}

The actual electron distribution in our PIC simulations (see Fig.\ref{fig5}a) cannot be accurately approximated by two counter-propagating maxwellians. However, the above analysis can be extended for a general electron distribution in the form of $f_{e0}(P_x, P_y)=\sum_{l=1}^N \delta(P_x-P_{dl})\exp(-P_y^2/2P_{tyl}^2)$, basically breaking the distribution into $N$ drifting beamlets. The thresholds in Eqs.\ref{threshold_f}-\ref{threshold_s} can be modified with the sum now extended to all $N$ beamlets. We have examined the electron distributions in two locations in our simulations, one in the shock region and the other in the target interior (shown by the small boxes in Fig.\ref{fig6}a), for their stability property. Specifically, we plot $\omega_{pl}^2(P_{dl}^2/P_{tyl}^2-1)$, which is basically  Eq.\ref{threshold_s} when $v_{txl}$ and $k$ are set to zero and $\omega_{pI}$ is neglected, for each beamlets in Fig.\ref{fig6}c and \ref{fig6}d. A positive dot contributes to the instability and a negative one to the stability. The sum of all the dots needs to be positive for the instability to occur. We found that the shock region is unstable (Fig.\ref{fig6}c) but the interior region is stable (Fig.\ref{fig6}d). Furthermore, if the space charge effect is included as in Eq.\ref{threshold_f} even the shock region becomes stable. Therefore, the electron distributions in these fast ignition simulations are marginally unstable due to their large angular spread.  The ions play an important role to neutralize any possible space charges. This is also supported by the observation that the ion density always display the same filament structure as the electrons. It is also worthwhile to point out that the return current beamlets (those with negative $P_{dl}$) contribute more than the forward beamlets toward the instability because of their relatively smaller angular spread. Any current filament stability analysis without considering the role of the return current would be inaccurate.

Due to dilution of the energetic electron relative density in the interior region, the electrons there are actually stable to the filament instability. There the magnetic pinching force cannot overcome the thermal pressure due to the large transverse momentum spread. The filament structure observed there is due to the electron streaming from the filaments in the shock region. It is therefore no surprise that no global coalescing of these filaments into a single strong filament is observed here (Fig.\ref{fig6}b), contrary to the previous simulations with smaller transverse box size\cite{pukhov:prl:fi,lasinski:pop:fi}, since the coalescing requires that the magnetic pinching force is larger than the thermal pressure.  Even without the global coalescing the magnetic field from the filament instability is still strong enough (reaching $\sim$300 MG in the shock region for $1\mu m$-laser) so that the gyro radius of MeV electrons is less than the laser wavelength. The electrons are confined by the self-generated magnetic field near the laser spot.

The difference regarding the filament coalescing embodies the difference in the electron distributions between different fast ignition simulations. In the simulations, the electron distribution can be affected by a number of factors. One factor is the total heat capacity of the target, which is determined by the target size and density used in the simulation. Given the amount of laser energy absorbed, the fraction of energetic electrons in the distribution will be determined by the heat capacity. The danger of using too small a target lies in that eventually all electrons in the simulation acquire MeV energy, a situation totally different from the actual fast ignition scheme. Another factor is the recycling of the energetic electrons, which is determined by the target size and the boundary conditions. If the transverse box size is small and no plasma-vacuum boundary is present, the electrons can recirculate through the laser spot many times transversely under the periodic boundary condition before they eventually leave in the longitudinal direction. The plasma-vacuum interface used here combined with the surface magnetic field prevents such transverse recirculation. However, the longitudinal recirculation is still present but occurs later than in a smaller sized target. Still another factor is the true 2D nature of the simulation afforded by the large transverse size used here. This allows the realistic simulation of the laser filaments and density channels which are important for the heating mechanisms\cite{res_heating_kru,brunel} which produce the large angular spread observed here.

\section{Summary and Discussion}
\label{IV}
In this paper the results from 2.5D PIC simulations for the interaction of a picosecond-long ignition laser pulse with a plasma pellet of 50-$\mu m$ diameter and 40-$n_c$ are presented. The high density pellet is surrounded by an underdense corona and is isolated by a vacuum region from the simulation box boundary.  The laser pulse is shown to filament and create density channels on the laser-plasma interface. The density channels increase the laser absorption efficiency and help generate an energetic electron distribution with a large angular spread. The combined distribution of the forward-going energetic electrons and the induced return electrons is marginally unstable to the current filament instability. The ions play an important role in neutralizing the space charges induced by the the temperature disparity between different electron groups. No global coalescing of the current filaments resulted from the instability is observed, consistent with the observed large angular spread of the energetic electrons.

The initial target $T_e$ used here is 7.4 keV, which is higher than the actual target bulk $T_e\approx 1$ keV. This $T_e$ is chosen mainly to avoid numerical self-heating\cite{birdsall}. Recently we have implemented a second-order-spline current deposition scheme in OSIRIS which when combined with the current smoothing allows a grid size of $\Delta x=12\lambda_D$ with virtually no self-heating. We repeat the simulations with $T_e=1.1$ keV and see no change in the main conclusions presented here. 

While the generation of the energetic electrons is studied here in considerable details, their transport from the birth place to $n=10^{23}/cm^3$ needs further study. The highly nonlinear nature of the interaction between the energetic electrons and emerging magnetic fields require simulations with targets of $\sim 100\mu m$-length and up to $100n_c$. The longitudinal recycling of the energetic electrons should also be eliminated through suitable boundary conditions. Only then can a reliable estimate of the energy flux reaching the $n=10^{23}/cm^3$ surface be obtained. Fortunately, with the advance of computer speed and PIC code improvement this type of simulations is within reach now.

For the very dense region of $n>10^{23}/cm^3$, the collisional effects become important and the electron transport and its final stopping will be affected by collisional scattering\cite{li-petrasso}. PIC or particle-fluid hybrid type of codes need to include a proper collision model to correctly simulate the physics there. Eventually the simulations need to be compared with experiments.

This work was supported by the US Department
of Energy through grants DE-FG02-03ER54721, DE-
FG02-03NA00065, and DE-FC02-04ER54789, by ILSA
at LLNL under W-07405-ENG48 and by NSF un-
der PHY032345. Simulations were done on the
DAWSON Cluster and at NERSC. The work of
Prof. L. O. Silva, Prof. R. A. Fonseca and
M. Fiore was partially supported by FCT (Portugal) through grants PDCT/FP/FAT/50190/2003, and
POCI/FIS/55905/2004.


\end{document}